\documentclass[preprint,showpacs,preprintnumbers,amsmath,amssymb]{revtex4}

\usepackage{graphicx}
\usepackage{dcolumn}
\usepackage{bm}

\begin{document}

\title{Non-Stationary Chaos}

\author{Taisei Kaizoji}%
\email{kaizoji@icu.ac.jp}
\affiliation{%
International Christian University, Mitaka,Tokyo 181-8585 Japan. 
}%


\date{\today}

\begin{abstract}
A new type of deterministic chaos for a system described by iterative two-dimensional maps is reported. The series being generated by the original map has an average upward trend while the first difference, which is the series of changes from one period to the next, exhibits chaotic behavior through period doubling bifurcation. Furthermore, step-wise time series appears as the values of the bifurcation parameter are large, and the first difference of the time series exhibits intermittent chaos. 

\end{abstract}


\maketitle

In 1974 Robert May [1,2] first reported on a one-dimensional map, such as the logistic map, which generates a chaotic dynamics. He proposed the non-linear difference equations as a model of population growth which takes place at discrete intervals of time and generations are completely non-overlapping. The time series appeared from the low-dimensional maps proposed by May is stationary. Namely, its statistical properties such as mean, variance, etc. are all constant over time. Therefore, theorems on existence of chaotic attractors generated by iterative maps [3,4] are necessarily based on the assumption that the time series can be rendered stationary. Since their studies, it has been believed that re-injection mechanisms to suppress exponential divergence of the variable are indispensable to emergence of chaos. \par
The aim of this letter is to show a new type of deterministic chaos for an iterative two-dimensional map which generates non-stationary time series. More concretely, the time series $ x(t) $ being appeared from the original map which we here propose has an average upward trend while the first difference of the time series of $ x(t) $, which is the series of changes from one period to the next, $ \Delta x(t) = x(t) - x(t-1) $ exhibits chaotic behavior through period doubling bifurcation. Furthermore, as the values of the bifurcation parameter are large, step-wise time series of $ x(t) $ appear and the first difference of $ x(t) $ exhibits intermittent chaos.  \par
Consider the two-dimensional difference equations

\begin{equation}
x(t+1) - x(t) = \alpha [\exp(y(t) - x(t)) - 1],
\end{equation}
\begin{equation}
y(t+1) - y(t) = \beta [x(t) - y(t)].
\end{equation}
Equation (2) can be transformed into an alternative form 
\begin{equation}
y(t) = \beta \sum^\infty_{i=1}(1 - \beta)^{i-1} x(t-i).
\end{equation}
The variable $ y(t) $ is an exponentially weighted moving average of $ x(t) $. The weight $ (1 - \beta)^{i-1} $ is assumed to be exponentially decreasing, that is, $ 0 < \beta < 1 $. After this, we specify the parameters as follows : $ \alpha > 0 $, and $ \beta = 0.5 $. Under these conditions, $ x(t) $ tends to go back to the exponentially weighted moving average, $ y(t) $. Namely, if $ y(t) > x(t) $, then $ x(t+1) $ increases, and $ y(t+1) $ decreases. The dynamic system (1)-(2) has infinite number of fixed points: $ x(t) = y(t) $. As can be checked easily, the local stability condition of a fixed point is $ \alpha + \beta < 2 $. 
The time series of $ x(t) $ has an average upward trend for the condition, $ \alpha + \beta > 2 $. The upward trend is robust with respect to the initial values of $ x(t) $ and $ y(t) $. Figure 1 is a sample path of $ x(t) $ as a function of time for $ \alpha = 5 $ and $ \beta = 0.5 $. The time series of $ x(t) $ fluctuates around the upward trend. To stationarize a time series of $ x(t) $, we take the first difference $ \Delta x(t) $ of a time series of $ x(t) $ which is the series of changes from one period to the next. Figure 2 shows that 
the corresponding time series of $ \Delta x(t) $ for $ \alpha = 5 $ and $ \beta = 0.5 $. The sample path of the first difference $ \Delta x(t) $ in Figure 2 is stationary and chaotic. 
To account for the chaotic behavior, let us consider the first return map of the form: 
\begin{equation}
\Delta x(t+1) = f(\Delta x(t)).
\end{equation}
Figure 3 is the bifurcation diagram for the first difference, $ \Delta x(t) $ generated by the first return map $ f(\Delta x(t)) $ where the parameter $ \alpha $ varies smoothly from 1 to 5. Figure 3 shows that chaos occurs through a sequence of period doubling as the parameter $ \alpha $ increases.
To construct the first return map, Figure 4 plots $ \Delta x(t) $ as a function of $ \Delta x(t) $ which is obtained for a chaotic regime of the attractor with $ \alpha = 5 $ and $ \beta = 0.5 $. 
One can reconstruct the graph of the first return map $ f(\Delta x(t)) $ from these observations. Figure 4 shows that we obtain a curve which has an extremum. 

The bifurcation diagram shows the period-three attractor located at $ \alpha = 6 $. Figure 5 show that 
the first retuen map $ f(\Delta x(t)) $  has a period-three point for $ \alpha \approx 2.9 $ and $ \beta = 0.5 $
The Li-York theorem [3] implies that such a map has points with every possible period. 

As $ \alpha $ increases further, step-wise time series of $ x(t) $ appears. Figure 6 is the time series of $ x(t) $ for $ \alpha = 15 $. Figure 7 is the first difference $ \Delta x(t) $ of the time series of $ x(t) $ in Figure 6. The time series of $ \Delta x(t) $ exhibits intermittency. As the parameter $\alpha $ is large, the laminar phase duration of intermittency is longer. \par
In this letter we propose a dynamical model which exhibits a new type of non-stationary chaos. Our model can be widely applied to non-stationary time series having an upward trend such as human population, World GDP, and so on. \par

\newpage
\begin{figure*}
\includegraphics{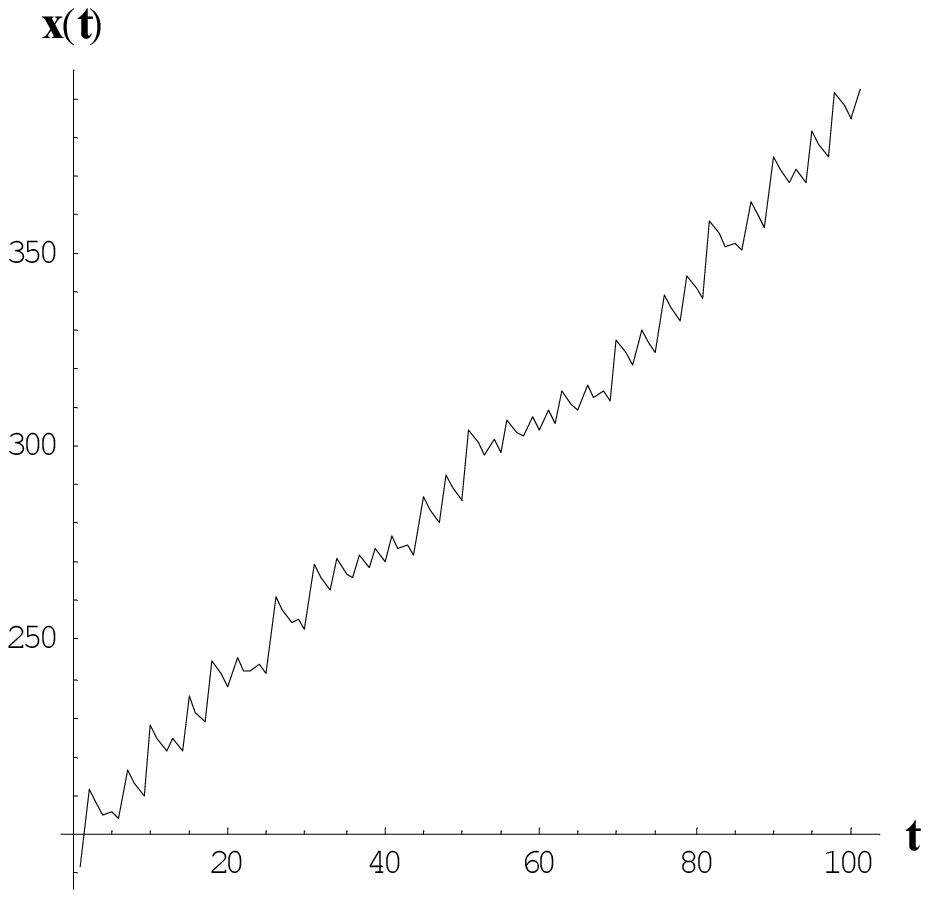}
\caption{The time series of $ x(t) $ as a function of time $ t $, as 
described by Equations (1) and (2) for $ \alpha = 4 $, and $ \beta = 0.5 $.}
\end{figure*}

\begin{figure*}
\includegraphics{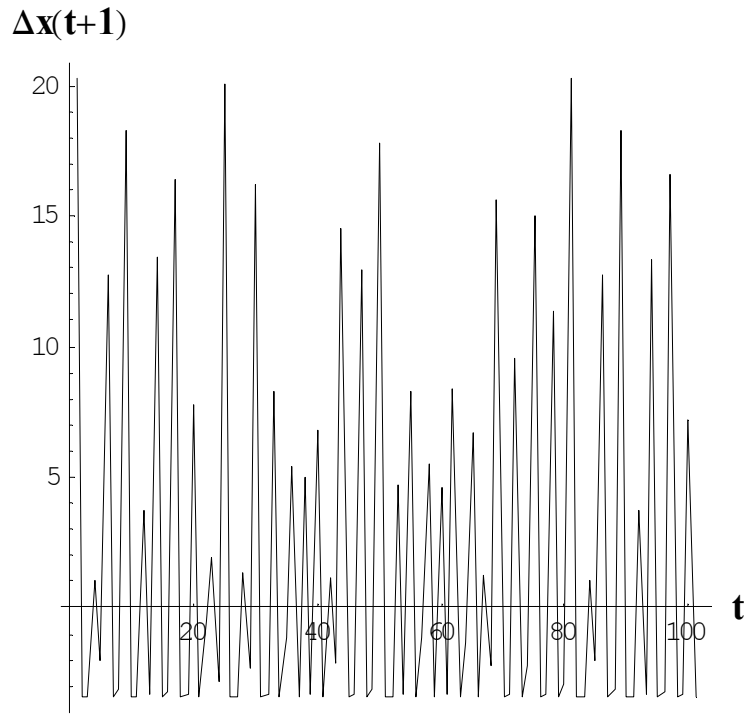}
\caption{The time series of $ \Delta x(t) = x(t+1) - x(t) $ as a function of time $ t $ for $ \alpha = 4 $, and $ \beta = 0.5 $. The time series corresponds to the first difference of the time series of $ x(t) $ in Figure 1.}
\end{figure*}

\begin{figure*}
\includegraphics{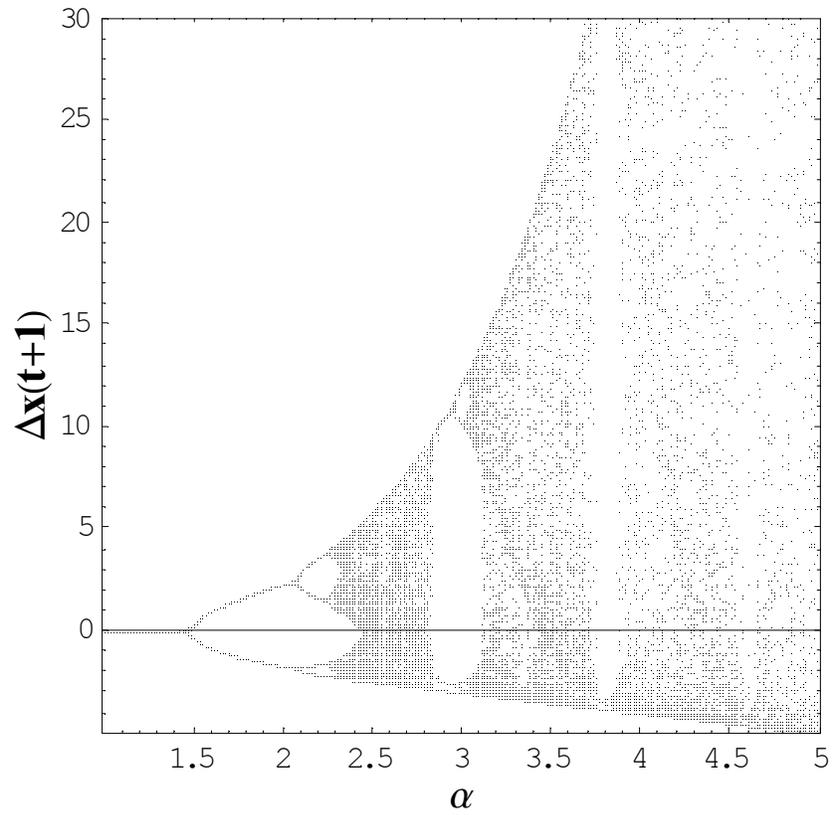}
\caption{The bifurcation diagram for the first difference of the time series of $ x(t) $, where the parameter $ \alpha $ varies smoothly from 1 to 5.}
\end{figure*}
\newpage
\begin{figure*}
\includegraphics{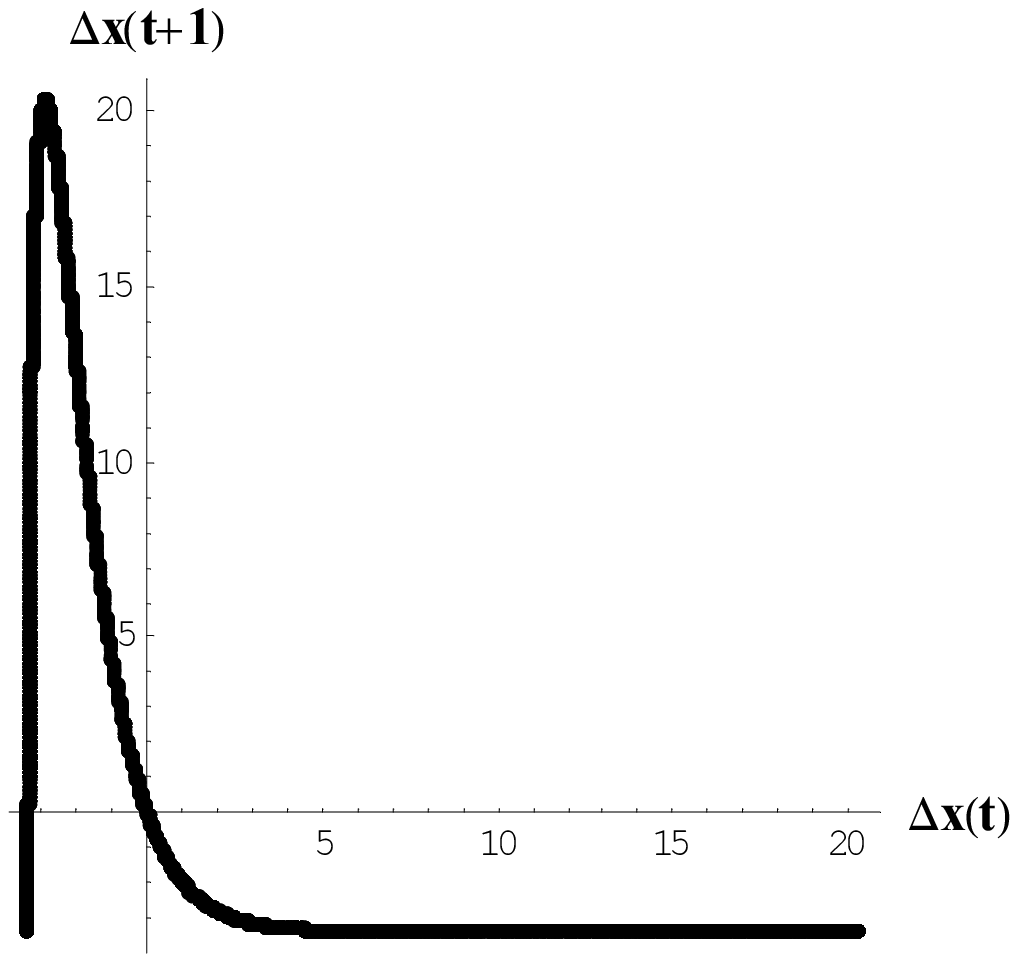}
\caption{The first return map $ f(\Delta x(t)) $ is plotted by using the time series of $ \Delta x(t) $ which is obtained for a chaotic regime of the attractor with $ \alpha = 5 $ and $ \beta = 0.5 $.}
\end{figure*}

\begin{figure*}
\includegraphics{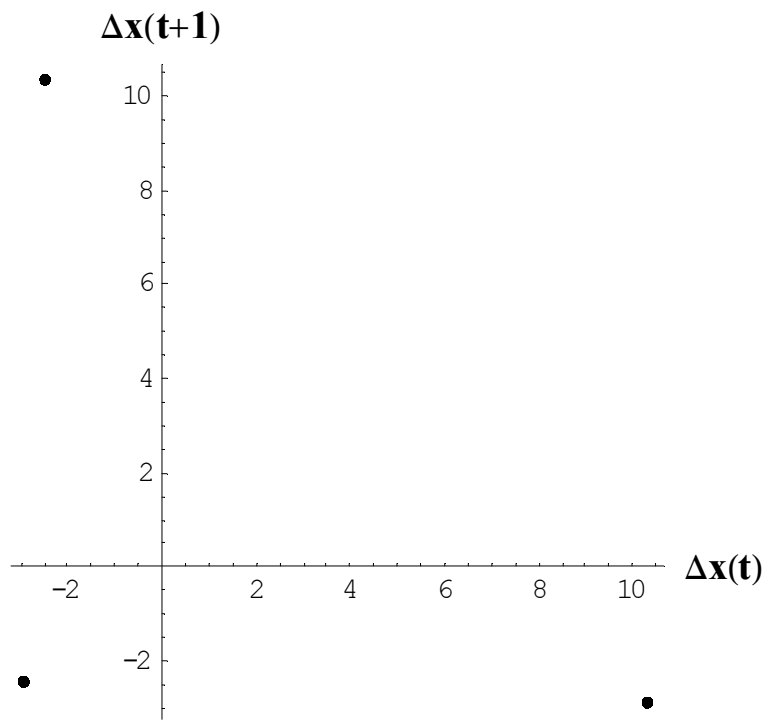}
\caption{The period-three of the first return map for $ \alpha = 2.9 $ and $ \beta = 0.5 $.}
\end{figure*}

\begin{figure*}
\includegraphics{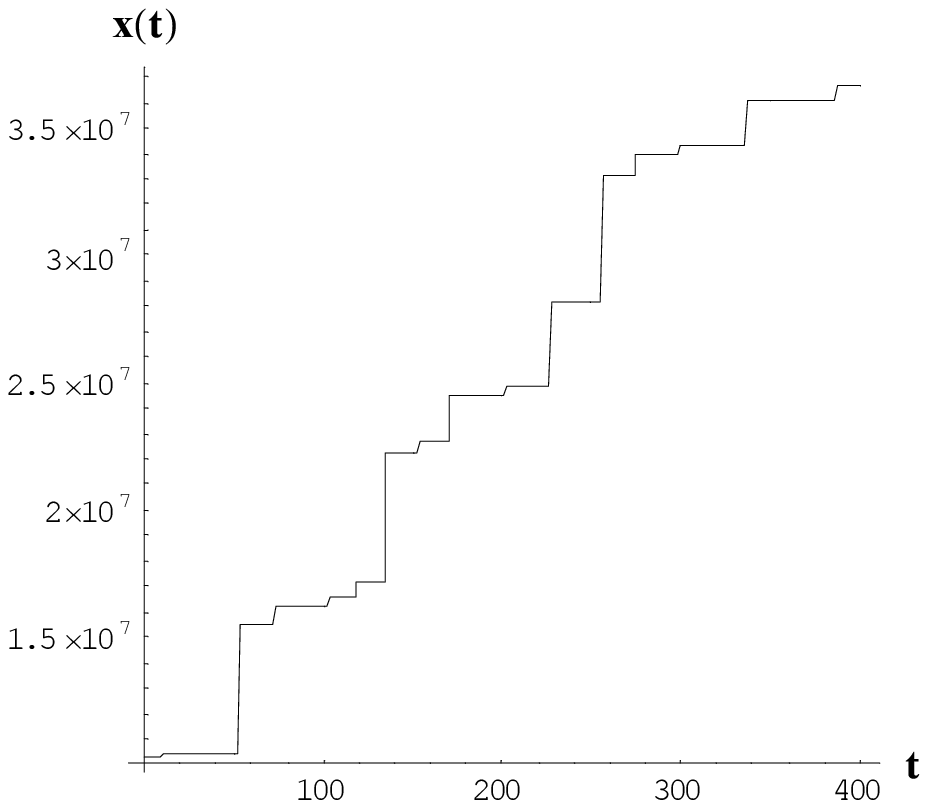}
\caption{The step-wise time series: The time series of $ x(t) $ for $ \alpha = 15 $, and $ \beta = 0.5 $. }
\end{figure*}

\begin{figure*}
\includegraphics{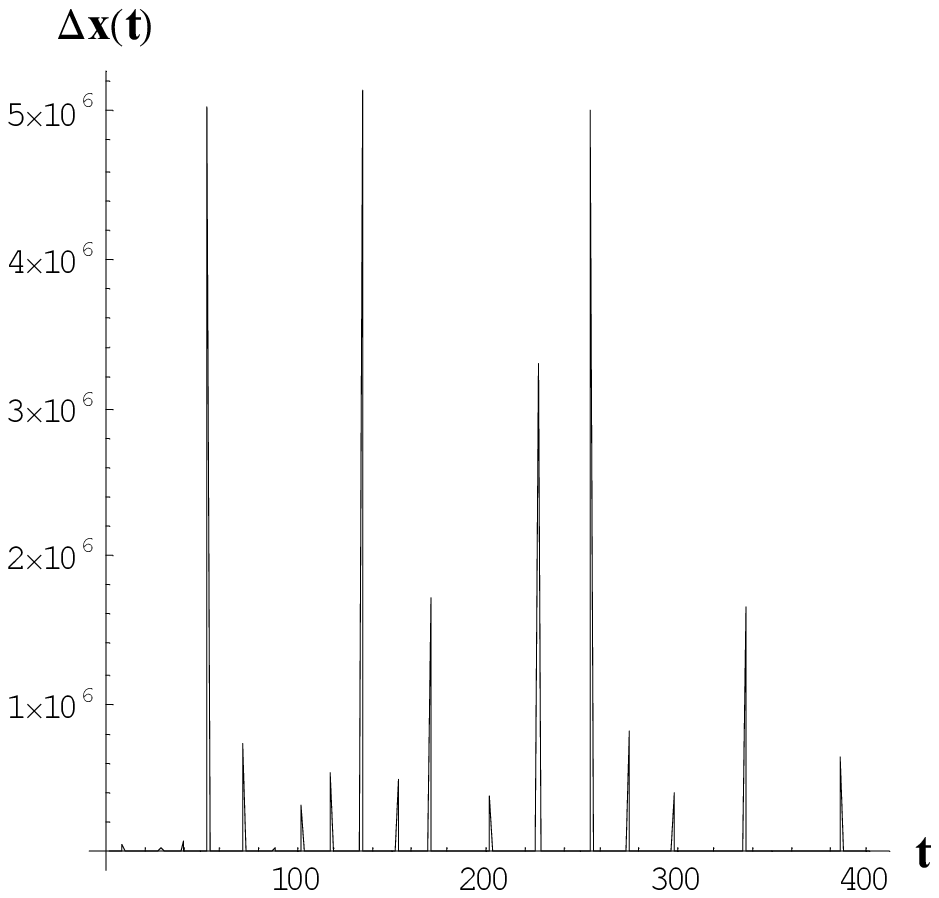}
\caption{Intermittency: The time series of $ \Delta x(t) = x(t+1) - x(t) $ for $ \alpha = 15 $, and $ \beta = 0.5 $. The time series corresponds to the first difference of the time series of $ x(t) $ in Figure 6.}
\end{figure*}

\end{document}